\documentclass[prb,aps,epsfig,twocolumn,showpacs]{revtex4}
\usepackage{color}
\usepackage{epsfig,amssymb,amsmath,latexsym}
\usepackage{subfigure}
\usepackage{wrapfig}
\usepackage{float}
\usepackage{color}
\usepackage{bm,bbm,dsfont,ulem,nicefrac}
\usepackage[colorlinks=true,linktoc=page,linkcolor=magenta,citecolor=magenta]{hyperref}

\newcommand\beq{\begin{equation}}
\newcommand\eeq{\end{equation}}

\newcommand\bea{\begin{eqnarray}}
\newcommand\eea{\end{eqnarray}}

\newcommand\bi{\begin{itemize}}
\newcommand\ei{\end{itemize}}

\newcommand\ie{{\it{i.e.}}}

\newcommand\smat{$\mathbb S$}

\newcommand\dc{{\textsf{DC~}}}

\newcommand\nw{{\textsf{NW~}}}
\newcommand\nwd{{\textsf{NW}}}
\newcommand\ff{{\textsf{FF~}}}
\newcommand\ffd{{\textsf{FF}}}
\newcommand\so{{\textsf{SOI~}}}
\newcommand\sod{{\textsf{SOI}}}

\newcommand\ffs{{\textsf{FFs~}}}
\newcommand\ffsd{{\textsf{FFs}}}

\newcommand\rnwqp{{\textsf{HRW~}}}
\newcommand\rnwqpd{{\textsf{HRW}}}
\newcommand\cdwqp{{\textsf{CDW~}}}
\newcommand\cdwqpd{{\textsf{CDW}}}

\newcommand\insbd{{\textsf{InSb}}}
\newif\ifboo \boofalse

\begin{document}

\textheight=23.8cm

\title{Quantum charge pumping through fractional Fermions in charge density modulated quantum wires and Rashba nanowires}%
\author{Arijit Saha, Diego Rainis, Rakesh P. Tiwari, and Daniel Loss}
\affiliation{
\mbox{Department of Physics, University of Basel, Klingelbergstrasse 82, CH-4056 Basel, Switzerland}\\
}

%-----------------------------------------
\date{\today}
%-------------------------------------------
%-------------------------------------------

\begin{abstract}
We study the phenomenon of adiabatic quantum charge pumping in systems supporting fractionally charged fermionic bound states, in two different setups.
The first quantum pump setup consists of a charge-density-modulated quantum wire, and the second one 
is based on a semiconducting nanowire with Rashba spin-orbit interaction, in the presence of a spatially oscillating magnetic field.
 In both these quantum pumps transport is investigated in a N-X-N geometry, with the system of interest (X) connected to two normal-metal leads (N), and the two pumping 
parameters are the strengths of the effective wire-lead barriers.
Pumped charge is calculated within the scattering matrix formalism. 
We show that quantum pumping in both setups provides a unique signature 
of the presence of the fractional-fermion bound states, in terms of asymptotically quantized pumped charge. 
Furthermore, we investigate shot noise arising due to quantum pumping,
verifying that quantized pumped charge corresponds to minimal shot noise.

\end{abstract}

\maketitle
%--------------------------------------------------------
\section{Introduction}
%--------------------------------------------------------
In recent years an exotic research line attracting considerable amount of attention has been focusing on condensed-matter systems where peculiar fractionally charged excitations emerge, 
which are interesting both from a fundamental point of view and for quantum computation purposes~\cite{ady1}. 
Some realizations of fractional fermions (\ffsd) in condensed-matter systems have already been proposed~\cite{gan12,kli12,sczhang}. These fractional fermion (\ffd)
bound states are predicted to be stable against weak disorder and interactions. The emergence of the \ffs in these systems can be understood by mapping the electronic 
low energy dynamics onto the Jackiw-Rebbi equations~\cite{jac76,rajaramanbell} describing massive Dirac fermions and the zero energy bound states of charge $e/2$ therein 
or onto the fractional charge formation in the Su$-$Schrieffer$-$Heeger model in long-chain polyenes~\cite{ssh1,ssh2}. 
These \ffs can also exhibit non-abelian braiding statistics~\cite{jelena3}.
Finally, it was shown that the presence of the \ffs could be revealed by transport experiments measuring two-terminal conductance, Aharonov-Bohm oscillations, 
and shot noise~\cite{die13}.

Adiabatic quantum pumping is a transport mechanism in meso- and nanoscale devices by which a finite \dc current is generated in the absence of an
applied bias by low-frequency periodic modulations of at least two system parameters~\cite{thouless,but94,bro98,bro2001}. 
The zero-bias current is obtained 
in response to the time variation of the parameters of the quantum system, which explicitly breaks time-reversal symmetry. 
Time-reversal symmetry breaking is necessary in order to get a pumped charge, but it is not a sufficient condition. 
Indeed, in order to obtain a finite net pumped charge, parity or spatial 
symmetry must also be broken. 
Finally, the required condition for electrical transport to be adiabatic consists in having a period $T$ of the oscillatory driving signals that has to be much longer 
than the dwell time $\tau_{\rm dwell}\simeq L/\upsilon_{\rm F}$ of the electrons inside the scattering region of length $L$, that is, $T = 2 \pi \omega^{-1} \gg \tau_{\rm dwell}$. 
In this limit, the pumped charge in a unit cycle becomes independent of the pumping frequency. This is referred to as ``adiabatic charge pumping''~\cite{bro98}.

In the last decades quantum charge and spin pumping through various mesoscopic setups, including quantum dots and quantum wires, has represented a fertile research line, 
both at the theoretical~\cite{niu1,niu2,spi95,shutenko,pb,b3,b4,b6,lev,ewaa1,ent,saha,aleiner,sharma2,andrei,sela,das2005,amit,cbenjamin,governale,tiw10,
gib13,pal13,Pekolaetal} and the experimental level~\cite{marcus,leek,buitelaar,giblin,blumenthal,Giazottoetal,Rocheelal,Connollyetal}, with focus on both the adiabatic and non-adiabatic regime.
The possible quantization of the charge pumped during a cycle through noninteracting open quantum systems~\cite{aleiner,lev,ewaa1,ent,saha,buitelaar}, as well as the circumstances 
under which the pump becomes ``optimal''~\cite{avr01,mos02}, are topics of fundamental interest.

Motivated by these works and by the recent advent of new exotic states of matter supporting peculiar bound states,
we study in this article adiabatic quantum charge pumping through
\ff bound states in two different configurations. 
Both the pumped charge and the noise obtained by adiabatic modulations of {at least} two system parameters can represent possible transport signatures for these \ffs other 
than conductance~\cite{die13}.

%---------------------------------------------------------------------------------------------------------------------
%---------------------------------------------------------------------------------------------------------------------
\begin{figure*}
%\begin{center}
\includegraphics[width=1.0\columnwidth]{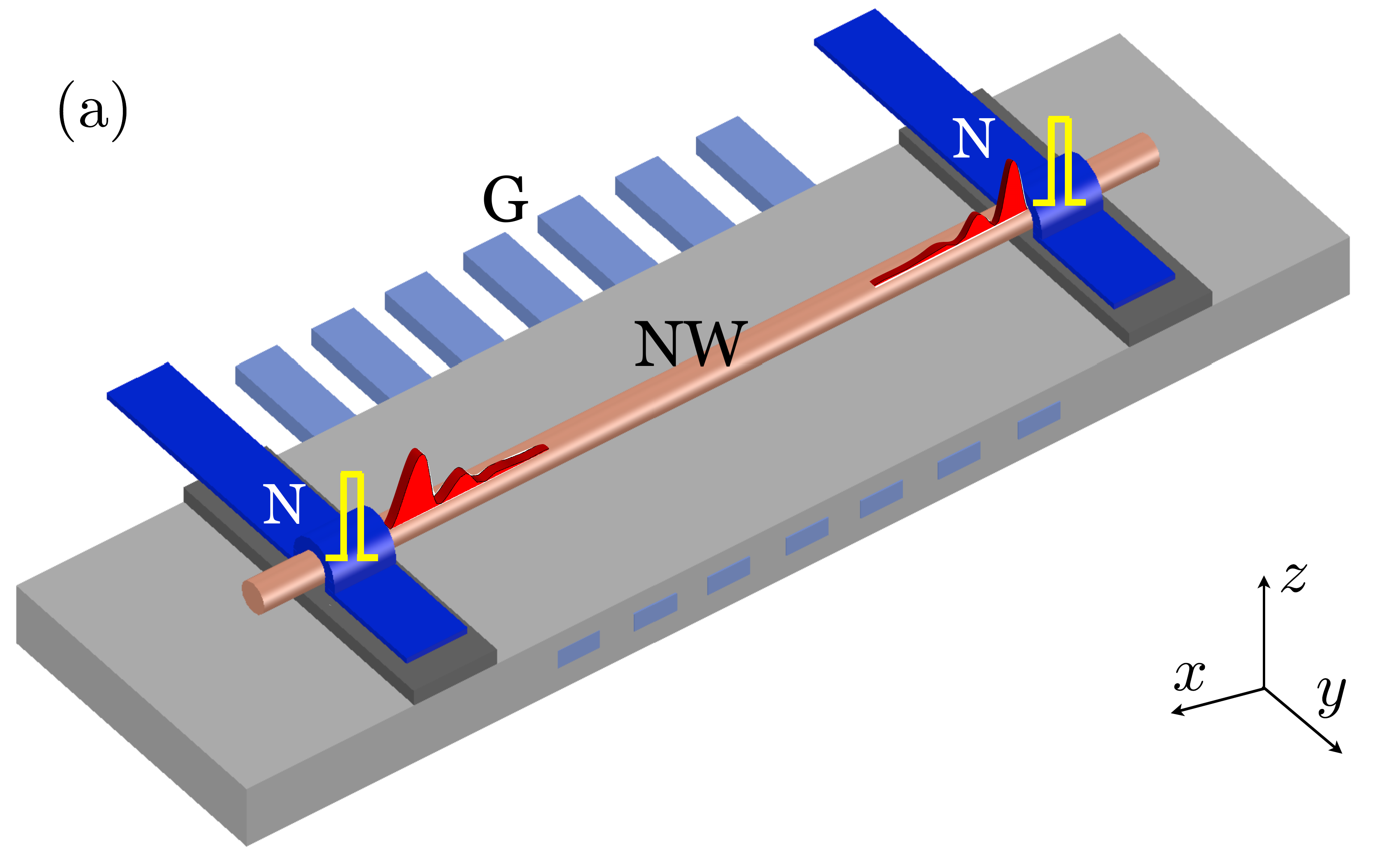}
\includegraphics[width=0.97\columnwidth]{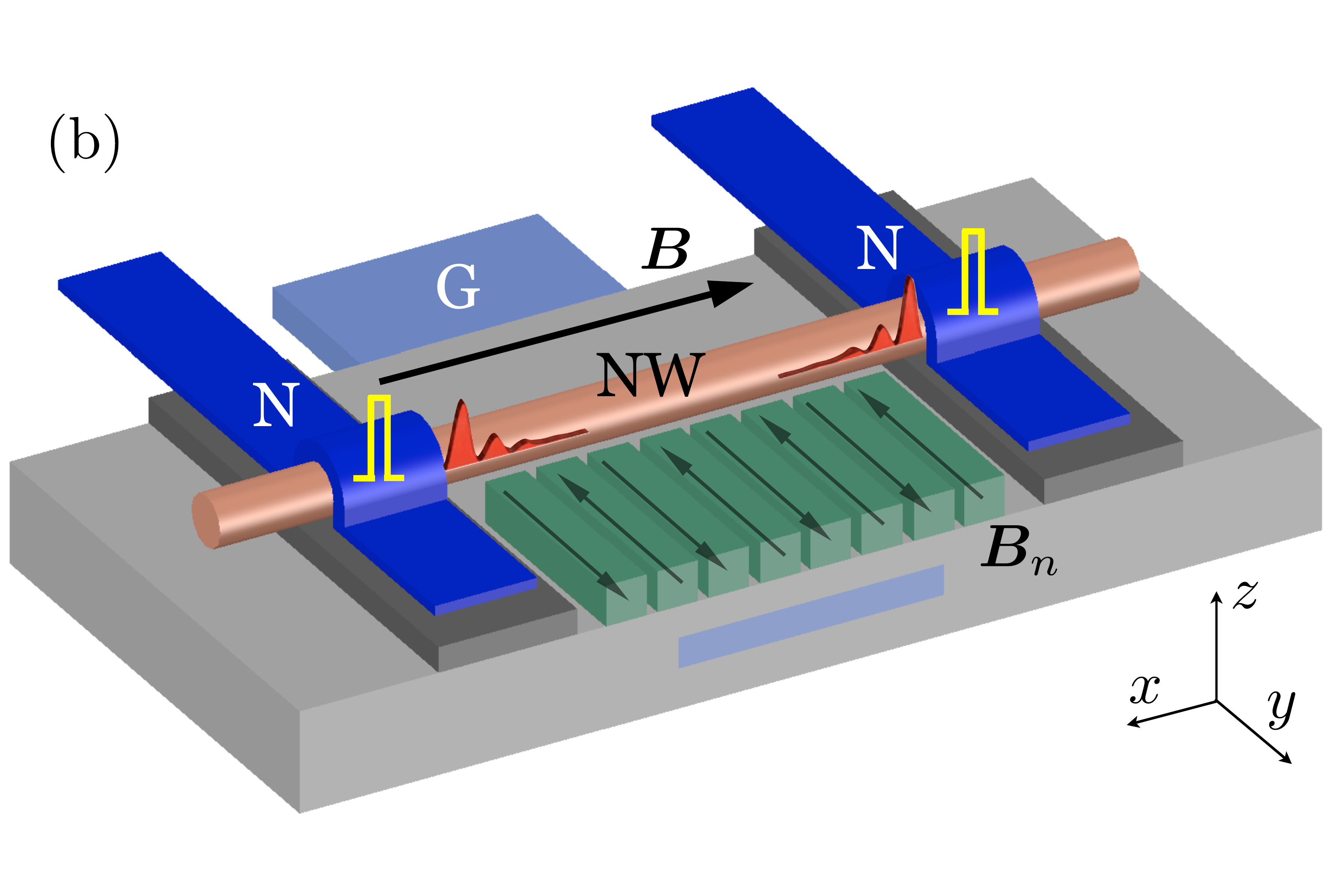}
\caption{(Color online) (a) Schematic of the charge-density-wave-based quantum pump in which the \nw (pink, light grey) of length $L$ is connected to two normal N leads (blue, black). 
In this geometry, the central \nw is subjected to charged gates G (light blue, light grey) which forms a charge-density-wave potential inside the \nwd. A \ff bound state emerges out
at each end of the wire (red, light grey). 
(b) Similar scheme for the helical-Rashba-nanowire-based quantum pump. The central part of the pump consists of a semiconducting \nw (pink, light grey) 
of length $L$ attached to two normal N leads (blue, black). A uniform magnetic field ${\bm {B}}$ is applied along the wire. The \nw is also subjected to a spatially varying magnetic 
field ${\bm {B}_{n}(x)}$ produced by periodically arranged nanomagnets (green, light grey). The gate G (light blue, light grey) controls the chemical potential in the \nwd. \ff bound states form 
at the two ends of the \nw (red, light grey). 
Two $\delta$-function barriers are symbolically denoted by the two yellow (light grey) rectangular barriers at each N-\nw interface in both cases. 
}
\label{fig1}
%\end{center}
\end{figure*}
%----------------------------------------------------------------------------------------------------------------------------------------
%----------------------------------------------------------------------------------------------------------------------------------------

We model our pump setups within the scattering matrix formalism~\cite{but94,bro98} and show that in both systems charge is pumped from one reservoir to the other via the \ffs present 
at the two ends of the nanowire. 
Thus, by measuring {current response} of these pumps one can demonstrate the existence of the \ffsd. 
Furthermore, we find that the shot noise in these pumps vanishes in correspondence to pumped charge being quantized, as expected. 
When the considered quantum pumps exhibit the above features, they are said to be {\it optimal}, with nearly quantized unit of charge being pumped in every cycle.

The remainder of this paper is organized as follows. 
In Sec.~\ref{sec:model}, we describe the two quantum pumps investigated here in detail, providing the linearized model Hamiltonians and the details of the pump mechanism. 
In Sec.~\ref{sec:transportquantities}, we present the expressions used to compute the pumped charge $\mathcal {Q}$ and the shot noise $\mathcal{S}_{\alpha\beta}$ for the two pumps 
within the scattering matrix framework. 
In Sec.~\ref{sec:results}, we present our numerical results for $\mathcal{Q}$ and $\mathcal{S}_{\alpha\beta}$ in these pump setups for various parameter regimes. 
Finally,  Sec.~\ref{sec:con} contains a summary of our numerical results followed by the conclusions.

%--------------------------------------------------------
\section{Model} \label{sec:model}
%--------------------------------------------------------
Here we introduce the two physical systems on which our pumping schemes are based. 
The first one is referred to as charge-density-wave wire (\cdwqpd) and the second one as helical Rashba nanowire (\rnwqpd).
In the \cdwqp case we consider spin-degenerate electrons. On the other hand, in the \rnwqp case we deal with spinful electrons.
As far as quantum pumping is concerned, 
the only difference between these two models lies in the fact that our results for the 
pumped charged in the \cdwqp refer to a single spin species and they have thus to be multiplied by two. 
%}}
%---------------------------------------------------------------------------------------------------------------------

%----------------------------------
\subsection{\cdwqpd}
%----------------------------------
The schematics of the \cdwqp is shown in Fig.~\ref{fig1}(a),
consisting of a nanowire (\nwd) of length $L$ with a gate-induced periodic potential, attached to two normal leads. 
The periodicity of the electrostatic potential is $\lambda_{\cdwqpd}=2\pi/k_{\cdwqpd}$~\cite{gan12}. 
The Hamiltonian describing the \nw is given by 
$H^{\rm \cdwqpd}= \int {\rm d}x \Psi^\dagger (x) \mathcal{H}^{\rm \cdwqpd} \Psi(x)$, where $\Psi(x)$ corresponds to the annihilation operator for an electron
at position $x$. 

The  Hamiltonian density for this spin-degenerate model reads

\begin{equation}
\mathcal{H}^{\rm \cdwqpd}=- \hbar^2 \partial_x ^2/2m-\mu + \Delta_{0}\cos (2 k_{\cdwqpd}x + \theta)
\label{eq:h_0_densitycdw}
\end{equation}
where $m$ is the effective mass of the electrons in the \nwd, $\mu$ the chemical potential and $\theta$ is a constant phase.

Assuming that the Fermi energy $m v_{\rm F}^2/2$ 
is the largest energy scale, following Ref.~\onlinecite{gan12} we linearize Eq.~(\ref{eq:h_0_densitycdw}) 
around the two Fermi points $k=\pm k_{\rm F}$. For $\mu=0$ we obtain the spectrum of the \nw at $k=\pm k_F$ as $E^2=(\hbar v_F k)^2 + \Delta_{0}^2$, with $v_F$ 
being the Fermi velocity. As explained in Ref.~\onlinecite{gan12}, this wire supports zero energy \ffs at the two ends of the \nw for $\theta=\pi/2$.

The Hamiltonian density describing the two normal non-interacting, spin-degenerate leads is 
$\mathcal{H}_l=-\hbar^2\partial_x^2/2m - \mu_l$, with $l$=L,R corresponding to 
left and right  lead, respectively, with chemical potential $\mu_l$.
The Fermi momentum is then $\hbar k_{l}=\sqrt{2m(\mu_l+E)}$. 
In this manuscript, as we are interested in quantum pumping we only consider the zero bias situation $\mu_{\rm L}=\mu_{\rm R}$.
We model the left and the right interface ($x=0$ and $x=L$) between the \nw and the normal leads by two  $\delta$-function barriers. 
The strengths of these $\delta$-function barriers can be controlled 
externally by applying additional gate voltages~\cite{giblin,Giazottoetal}, which could be different at the left and the right interfaces. 
In our quantum pump the two pump parameters are these left and right $\delta$-function barrier strengths, evolving in time either as (off-set circular contours)
\begin{align}\label{circular}
\lambda_1&=\lambda_0+P_s \cos(\omega t - \phi)\; \nonumber \\
\lambda_2&=\lambda_0+P_s \cos(\omega t + \phi)
\end{align}
or as (``lemniscate'' contours)
\begin{align}\label{lemniscate}
\lambda_1&=P_s (\cos\Theta\cos\omega t - \sin\Theta\sin\omega t\cos\omega t)/(1 + \sin\omega t)^2\;  \nonumber  \\
\lambda_2&=P_s (\cos\Theta\cos\omega t + \sin\Theta\sin\omega t\cos\omega t)/(1 + \sin\omega t)^2\;,
\end{align}
 respectively.
In the circular contour, $\lambda_{0}$ is  the mean value of the amplitude around which the two pumping parameters are modulated with time.
In both cases $P_s$ is called the pumping strength. 
Further, $2\phi$ and $\Theta$ are the phase offsets between the two 
pumping signals for the circular and lemniscate contours respectively. 
Such parametric curves in the $\lambda_1$--$\lambda_2$ plane are shown in Figs.~\ref{fig2} and \ref{fig4}.

In our analysis we consider only adiabatic quantum pumping, valid in the regime where the time period of the pump parameters $T=2\pi\omega^{-1}$ 
is much larger than the dwell time $\tau_{\rm dwell}\simeq L/v_F$ of the electrons inside the \nw,  \ie,~$T\gg\tau_{\rm dwell}$.

\vspace{-0.3cm}
%---------------------------------
\subsection{\rnwqpd}
%---------------------------------
In Fig.~\ref{fig1}(b) we show the schematics of the \rnwqpd, consisting of a Rashba nanowire attached to two normal leads. 
The central part of this pump consists of a semiconducting wire of length $L$ along the $\hat{x}$ direction with a finite Rashba spin orbit interaction (\sod) 
and an external magnetic field, which has both a uniform ($\textbf{B}$) and a spatially varying ($\textbf{B}_n$) component. 
The corresponding Hamiltonian describing this \nw is given by
$H^{\rnwqpd}=  \int {\rm d}x \Psi^\dagger (x) \mathcal{H}^{\rnwqpd} \Psi (x)$, where 
 $\Psi = (\Psi_\uparrow, \Psi_\downarrow)$ with $\Psi_{\sigma}(x)$
being the annihilation operator for a spin $\sigma$ ($\in\{\uparrow,\downarrow\})$ electron at position $x$. 

The Hamiltonian density for this spinful model is given by
\begin{equation}
\mathcal{H}^{\rnwqpd}=- \hbar^2 \partial_x ^2/2m-\mu - i \alpha  \sigma_z  \partial_x  + \frac{g\mu_B}{2}  [{\bm B}+{\bm B}_{n}(x)]  \cdot {\boldsymbol \sigma}   \;.
\label{eq:h_0_density}
\end{equation}
Here $m$ is the effective electron mass in the \nwd, $\mu$ the chemical potential, $\alpha$ the \so coefficient and $\sigma_i$ the usual Pauli spin matrices. 
Further, $g$ is the Lande g-factor and $\mu_B$ the Bohr magneton. 
We choose the uniform field $\bm B$ to be pointing along the $\hat x$ direction,
opening up a Zeeman gap of magnitude $\Delta_z=g \mu_B B$ at $k=0$.  
The spatially periodic magnetic field $\bm B_n$ is oriented along 
the $\hat y$ direction, ${\bm B}_{n,x}={\hat y}B_{n}\sin (4 k_{\rm so} x+\theta)$, couples the two exterior branches of the spectrum~\cite{kli12,die13} 
and opens up a gap of magnitude $\Delta_n=g \mu_B B_n/2$ at $k=\pm 2k_{\rm so}$.  
Assuming that the \so energy $m\alpha^2/2\hbar^2$ is 
the largest energy scale at the chemical potential, following Ref.~[\onlinecite{jelena4,kli12,die13}], 
we can linearize the Hamiltonian $\mathcal{H}^{\rnwqpd}$ around $k=0$ (interior branches) and $k=\pm k_{\rm so}$ (exterior branches).
For $\mu=0$ one obtains the spectrum of the \nw around $k=0$ and $k=\pm2k_{\rm so}$ as
$E^2=(\hbar \upsilon_F k)^2+\Delta_z^2 \ $ and $E^2=(\hbar \upsilon_F k)^2+\Delta_n^2 \ $ respectively, with Fermi velocity 
$v_{\rm F}=\alpha/\hbar$.
As shown in Ref.~[\onlinecite{kli12,die13}] this system is fully gapped and supports \ff bound states localized at the two ends of the \nwd, with degenerate 
zero energy for $\theta=\pi$.

Like in  the \cdwqp case, the Hamiltonian density for the two normal non-interacting, spin-degenerate leads is 
$\mathcal{H}_l=-\hbar^2\partial_x^2/2m - \mu_l$, where $l$=L,R, and $\mu_l$ denotes the corresponding chemical potential ($\mu_{\rm L}=\mu_{\rm R}$), with Fermi momentum 
$\hbar k_{l}=\sqrt{2m(\mu_l+E)}$. 
Again, the left and the right interfaces ($x=0$ and $x=L$) between the \nw and the normal leads are modeled by two different $\delta$-function barriers,  
 whose heights represent the two pumping parameters, and evolve according to the two possible paths given by Eqs.~(\ref{circular}) and (\ref{lemniscate}).
As before, the pumping strength is denoted by $P_s$, while $2\phi$ (circular) and $\Theta$ (lemniscate) are the phase difference between the two pumping parameters. 
Again, we restrict our analysis to the adiabatic quantum pumping regime $T\gg L/v_F$.

%---------------------------------------------------------------------
\section{Pumped charge and noise} \label{sec:transportquantities}
%---------------------------------------------------------------------
To calculate the pumped charge we use Brouwer's formula~\cite{bro98}, which relies on the knowledge of the \smat-matrix for the two systems considered here. 
The shot noise due to the pump can also be expressed in terms of the \smat-matrix elements, as done for example in Ref.~\onlinecite{mos02}.

%--------------------------------------------------------
\subsection{\cdwqpd}
%--------------------------------------------------------
The general 2$\times$2 \smat-matrix for the \cdwqp geometry can be written as
\bea
\mathbb{S}_{\cdwqp} =
\begin{bmatrix}~|r|e^{i\gamma} & |t|e^{i\psi}~\\
~|t^{\prime}|e^{i\psi^{\prime}} 
& |r^{\prime}|e^{i\gamma^{\prime}}~\\
\end{bmatrix}
.
\label{smatcdw}
\eea
We write here the complex \smat-matrix elements \smat$_{ij}$ in polar form, with modulus and phase explicitly shown, since the phase is going to play a major role in the determination of the pumped charge.
The \smat$_{ij}$ are all functions of the incident energy $E$ 
and depend parametrically  on the nanowire length $L$, the \cdwqp gap $\Delta_{0}$, the phase $\theta$ associated to the charge density wave, 
and the strengths $\lambda_1$, $\lambda_2$ of the two $\delta$-function barriers at $x=0$ and $x=L$, respectively.

%-----------------------------------
\subsubsection{Pumped charge}
%-----------------------------------
Following Ref.~\onlinecite{bro98}, for an electron incident from the left lead (L), the  formula for the pumped charge can be obtained from the parametric derivatives of the 
$\mathbb{S}$-matrix elements. For the spinless, single-channel case considered here, one has
\begin{eqnarray}
\mathcal {Q}_{\cdwqp} &=& \frac{e}{2 \pi} \int\limits_{0}^{\tau} dt \Big[|r|^{2}\dot{\gamma} + |t|^{2}\dot{\psi}\Big]\ .
\label{pump1cdw}
\end{eqnarray}

%-----------------------------
\subsubsection{Shot Noise}
%-----------------------------
The noise properties of the \cdwqp are investigated within the scattering matrix formalism for ac transport~\cite{mos02}. In general, the current-current correlation function is expressed as
\begin{equation}
\mathcal{S}_{\alpha\beta}(t,t^{\prime})=\frac{1}{2}\langle\Delta \hat{I}_\alpha (t) \Delta \hat{I}_\beta (t^{\prime})
+\Delta \hat{I}_\beta (t^{\prime})\Delta \hat{I}_\alpha (t)\rangle,
\label{eq:cccor}
\end{equation}
depending on two time instants $t$ and $t^{\prime}$, with $\Delta \hat{I}=\hat{I}-\langle\hat{I}\rangle$ and $\hat{I}_\alpha(t)$ being the quantum-mechanical current operator 
in lead $\alpha$. Since we are interested only in correlations over long time intervals ($|t^\prime- t|\gg 2\pi/\omega$), we investigate 

\begin{equation}
\mathcal{S}_{\alpha\beta}(t)=\frac{\omega}{2\pi}\int_0^{2\pi/\omega} dt^{\prime} \mathcal{S}_{\alpha\beta} (t,t^{\prime}).
\label{eq:noiseaverage}
\end{equation}

Let us introduce $\mathcal{S}_{\alpha\beta}^{\text{pump}}$ as the zero-frequency component of the above long-time averaged correlator,
$\mathcal{S}_{\alpha\beta}^{\text{pump}}=\int dt\mathcal{S}_{\alpha\beta}(t)$. In the low-temperature limit the 
noise power produced by the pump can be separated into two parts~\cite{mos02}
\begin{equation}
\mathcal{S}_{\alpha\beta}^{\text{pump}}=\delta_{\alpha\beta}\mathcal{S}_\alpha^{\text{pump,P}} + 
\mathcal{S}_{\alpha\beta}^{\text{pump,cor}},
\label{eq:noisesplit}
\end{equation}
where the first part $\mathcal{S}_\alpha^{\text{pump,P}}$ is due to an uncorrelated motion of nonequilibrium quasielectrons and quasiholes (Poissonian  component). 
The second part $\mathcal{S}_{\alpha\beta}^{\text{pump,cor}}$ denotes the contribution from correlations between the quasielectrons and quasiholes. 
These correlations correspond to processes where 
first a quasielectron-quasihole pair is created by absorption of an energy quantum $\hbar\omega$ and then the quasielectron and the quasihole belonging to the same pair get scattered into 
different leads. As we are interested in the noise generated by two simultaneously oscillating parameters $\lambda_1$ and $\lambda_2$, we must differentiate between the 
noise produced by the variation of each single parameter $\lambda_1$ or $\lambda_2$ separately, and the additional noise generated by  quantum pumping. The latter is denoted by

\begin{align}
\Delta\mathcal{S}_{\alpha\beta}^{\text{pump}}=\delta_{\alpha\beta}\Delta\mathcal{S}_\alpha^{\text{pump,P}}+\Delta\mathcal{S}_{\alpha\beta}^{\text{pump,cor}}, 
\label{eq:extranoise}
\end{align}
where 
$\delta_{\alpha\beta}\Delta\mathcal{S}_\alpha^{\text{pump,P}}$ and $\Delta\mathcal{S}_{\alpha\beta}^{\text{pump,cor}}$ denote the contribution to the additional noise coming 
from the first and the second term on the right hand side of Eq.~(\ref{eq:noisesplit}) respectively.

We obtain the following expression for such additional noise:
\begin{align}
\Delta\mathcal{S}_{\rm L}^{\text{pump,P}}=\frac{e^2\omega}{\pi}\cos(2\phi)\Bigg[ &
|\dot{r}|^2+|r|^2\dot{\gamma}^2 \nonumber \\
+ & |\dot{t}|^2+|t|^2\dot{\psi}^2 \Bigg] \ ,
\label{eq:deltaslcdw}
\end{align}
which is integrated over the pumping contour following Eq.~(\ref{eq:noiseaverage}). Similarly, 
\begin{eqnarray}
\Delta\mathcal{S}_{\rm LL}^{\text{pump,cor}}&=&-\frac{e^2\omega}{\pi}\cos(2\phi)\Bigg[
(|\dot{r}|^2+|r|^2\dot{\gamma}^2 + |\dot{t}|^2 \nonumber \\ 
&+&|t|^2\dot{\psi}^2) +\Big(|t|^2(|\dot{r}|^2+|r|^2\dot{\gamma}^2) + |r|^2(|\dot{t}|^2\nonumber \\ 
&+&|t|^2\dot{\psi}^2) -2|r||t|(|\dot{r}||\dot{t}|+|r||t|\dot{\gamma}\dot{\psi})\Big) \Bigg].
\label{eq:deltasllcdw}
\end{eqnarray}
Adding Eq.~(\ref{eq:deltaslcdw}) and Eq.~(\ref{eq:deltasllcdw}) we obtain the auto-correlator $\Delta\mathcal{S}_{\rm LL}^{\text{pump}}$ for the \cdwqpd. 

In general, the expression for the additional noise depends on both the time derivatives of the scattering amplitudes and their phases. The Poissonian
part of the additional pump noise given by Eq.~(\ref{eq:deltaslcdw}), contains only even powers of the phase derivatives while $\Delta\mathcal{S}_{\rm LL}^{\text{pump,cor}}$
contains bilinear terms involving time derivatives of the phases for both the reflection and transmission amplitude. The effect of the latter can be characterized in terms of 
the Fano factor as defined in the next section.

\vspace {1cm}
%
%--------------------------------------------------------
\subsection{\rnwqpd}
%--------------------------------------------------------
The most general 4$\times$4 \smat-matrix for the \rnwqp geometry can be expressed as (the matrix elements are given in polar form, similarly to the \cdwqp case)
\bea
\mathbb{S}_{\rnwqp} =
\begin{bmatrix}~|r_{\uparrow\uparrow}|e^{i\gamma} & |r_{\uparrow\downarrow}|e^{i\delta} & |t_{\uparrow\uparrow}|e^{i\psi} & |t_{\uparrow\downarrow}|e^{i\eta}~\\
~|r_{\downarrow\uparrow}|e^{i\tilde{\delta}} & |r_{\downarrow\downarrow}|e^{i\tilde{\gamma}} & |t_{\downarrow\uparrow}|e^{i\tilde{\eta}} & |t_{\downarrow\downarrow}|e^{i\tilde{\psi}}~\\
~|t^{\prime}_{\uparrow\uparrow}|e^{i\psi^{\prime}} & |t^{\prime}_{\uparrow\downarrow}|e^{i\eta^{\prime}} & |r^{\prime}_{\uparrow\uparrow}|e^{i\gamma^{\prime}} 
& |r^{\prime}_{\uparrow\downarrow}|e^{i\delta^{\prime}}~\\
~|t^{\prime}_{\downarrow\uparrow}|e^{i\tilde{\eta}^{\prime}} & |t^{\prime}_{\downarrow\downarrow}|e^{i\tilde{\psi}^{\prime}} 
& |r^{\prime}_{\downarrow\uparrow}|e^{i\tilde{\delta}^{\prime}} & |r^\prime_{\downarrow\downarrow}|e^{i\tilde{\gamma}^{\prime}}~\\
\end{bmatrix}
.
\label{smatnw}
\eea
Also here, the \smat-matrix elements 
are functions of the incident energy $E$ and depend on the wire length $L$, the two Zeeman gaps $\Delta_{z}$, $\Delta_{n}$, the phase $\theta$ 
associated to the spiral magnetic field, and the strengths $\lambda_1$, $\lambda_2$ of the two lead-wire barriers.
%------------------------------------------------------------------------------------------------------------------------

%-----------------------------------
\subsubsection{Pumped charge}
%-----------------------------------
Following Ref.~[\onlinecite{bro98}], the formula for the pumped charge for this spinful single-channel wire reads
\begin{align}
\mathcal {Q}_{\rnwqp} = \frac{e}{2 \pi} \int\limits_{0}^{\tau} dt \Big[ & |r_{\uparrow\uparrow}|^{2}\dot{\gamma} + |r_{\uparrow\downarrow}|^{2}\dot{\delta} \nonumber \\
+  & |t_{\uparrow\uparrow}|^{2}\dot{\psi} + |t_{\uparrow\downarrow}|^{2}\dot{\eta}\Big]\ . 
\label{pump1}
\end{align}

%--------------------------------------
\subsubsection{Shot Noise}
%--------------------------------------

Following arguments similar to the case of \cdwqp we obtain the spinful noise expressions for the \rnwqp case as
%\begin{widetext}
\begin{eqnarray}
\Delta\mathcal{S}_{\rm L}^{\text{pump,P}}&=&\frac{e^2\omega}{\pi}\cos(2\phi)\Bigg[
|\dot{r}_{\uparrow\uparrow}|^2+|r_{\uparrow\uparrow}|^2\dot{\gamma}^2 \nonumber \\
&+& |\dot{r}_{\uparrow\downarrow}|^2+|r_{\uparrow\downarrow}|^2\dot{\delta}^2 +
|\dot{t}_{\uparrow\uparrow}|^2 \nonumber \\
&+&|t_{\uparrow\uparrow}|^2\dot{\psi}^2 +
|\dot{t}_{\uparrow\downarrow}|^2+|t_{\uparrow\downarrow}|^2\dot{\eta}^2  
\Bigg],
\label{eq:deltasl}
\end{eqnarray}
%\end{widetext}
which is integrated over the pumping contour following Eq.~(\ref{eq:noiseaverage}). For the additional noise, this gives us the contribution due to the uncorrelated motion of 
quasielectrons and quasiholes in the same lead L (auto-correlator), corresponding to the first term on the right hand side of Eq.~(\ref{eq:extranoise}). 
Similarly, for the second part of the auto-correlator we obtain

\begin{widetext}
\begin{eqnarray}
\Delta\mathcal{S}_{\rm LL}^{\text{pump,cor}}&=&-\frac{e^2\omega}{\pi}\cos(2\phi)\Bigg[
|r_{\uparrow\uparrow}|^2(|\dot{r}_{\uparrow\uparrow}|^2+|r_{\uparrow\uparrow}|^2\dot{\gamma}^2) +
2|r_{\uparrow\uparrow}||r_{\uparrow\downarrow}|(|\dot{r}_{\uparrow\uparrow}||\dot{r}_{\uparrow\downarrow}|+ |r_{\uparrow\uparrow}||r_{\uparrow\downarrow}|\dot{\gamma}\dot{\delta}) \nonumber \\
&+&
2|r_{\uparrow\uparrow}||t_{\uparrow\uparrow}|(|\dot{r}_{\uparrow\uparrow}||\dot{t}_{\uparrow\uparrow}|+ |r_{\uparrow\uparrow}||t_{\uparrow\uparrow}|\dot{\gamma}\dot{\psi}) +
2|r_{\uparrow\uparrow}||t_{\uparrow\downarrow}|(|\dot{r}_{\uparrow\uparrow}||\dot{t}_{\uparrow\downarrow}|+ |r_{\uparrow\uparrow}||t_{\uparrow\downarrow}|\dot{\gamma}\dot{\eta}) +
|r_{\uparrow\downarrow}|^2(|\dot{r}_{\uparrow\downarrow}|^2+ |r_{\uparrow\downarrow}|^2\dot{\delta}^2) \nonumber \\
&+&
2|r_{\uparrow\downarrow}||t_{\uparrow\uparrow}|(|\dot{r}_{\uparrow\downarrow}||\dot{t}_{\uparrow\uparrow}|+ |r_{\uparrow\downarrow}||t_{\uparrow\uparrow}|\dot{\delta}\dot{\psi}) +
2|r_{\uparrow\downarrow}||t_{\uparrow\downarrow}|(|\dot{r}_{\uparrow\downarrow}||\dot{t}_{\uparrow\downarrow}|+ |r_{\uparrow\downarrow}||t_{\uparrow\downarrow}|\dot{\delta}\dot{\eta}) +
|t_{\uparrow\uparrow}|^2(|\dot{t}_{\uparrow\uparrow}|^2+ |t_{\uparrow\uparrow}|^2\dot{\psi}^2) \nonumber \\
&+&
2|t_{\uparrow\uparrow}||t_{\uparrow\downarrow}|(|\dot{t}_{\uparrow\uparrow}||\dot{t}_{\uparrow\downarrow}|+ |t_{\uparrow\uparrow}||t_{\uparrow\downarrow}|\dot{\psi}\dot{\eta}) + 
|t_{\uparrow\downarrow}|^2(|\dot{t}_{\uparrow\downarrow}|^2+ |t_{\uparrow\downarrow}|^2\dot{\eta}^2)
\Bigg].
\label{eq:deltasll}
\end{eqnarray}
\end{widetext}

Adding Eq.~(\ref{eq:deltasl}) and Eq.~(\ref{eq:deltasll}) we obtain the spinful auto-correlator ($\Delta\mathcal{S}_{\rm LL}^{\text{pump}}$) for the \rnwqp case.

%---------------------------------------------------------------------------------------------------------------
\section{Numerical results} \label{sec:results}
%---------------------------------------------------------------------------------------------------------------
In this section we present and discuss the outcome of our numerical results for the pumped charge and noise through the \ff bound states in case of \cdwqp 
and the \rnwqpd, respectively. For the \cdwqp pump, the Fermi energy $m v_{\rm F}^2/2$ is the largest energy scale in our analysis. Hence, the other 
energy scales ($\Delta_0, \omega$) are expressed in terms of $v_F$. 
On the other hand the largest energy scale in our analysis  for the \rnwqp is the \so energy $m\alpha^2/2\hbar^2$, and  the other energy scales ($\Delta_z,\Delta_n,\omega$) 
are then expressed in terms of $\alpha$. Also, throughout our analysis we have set $m=1$ and $\hbar=1$.

%====================================================================
%====================================================================
\subsection{\cdwqpd}
%====================================================================
%====================================================================

%---------------------------------------------------------------------------------------------------------------
\begin{figure}[h!]
\begin{center}
\includegraphics[width=1.0\linewidth]{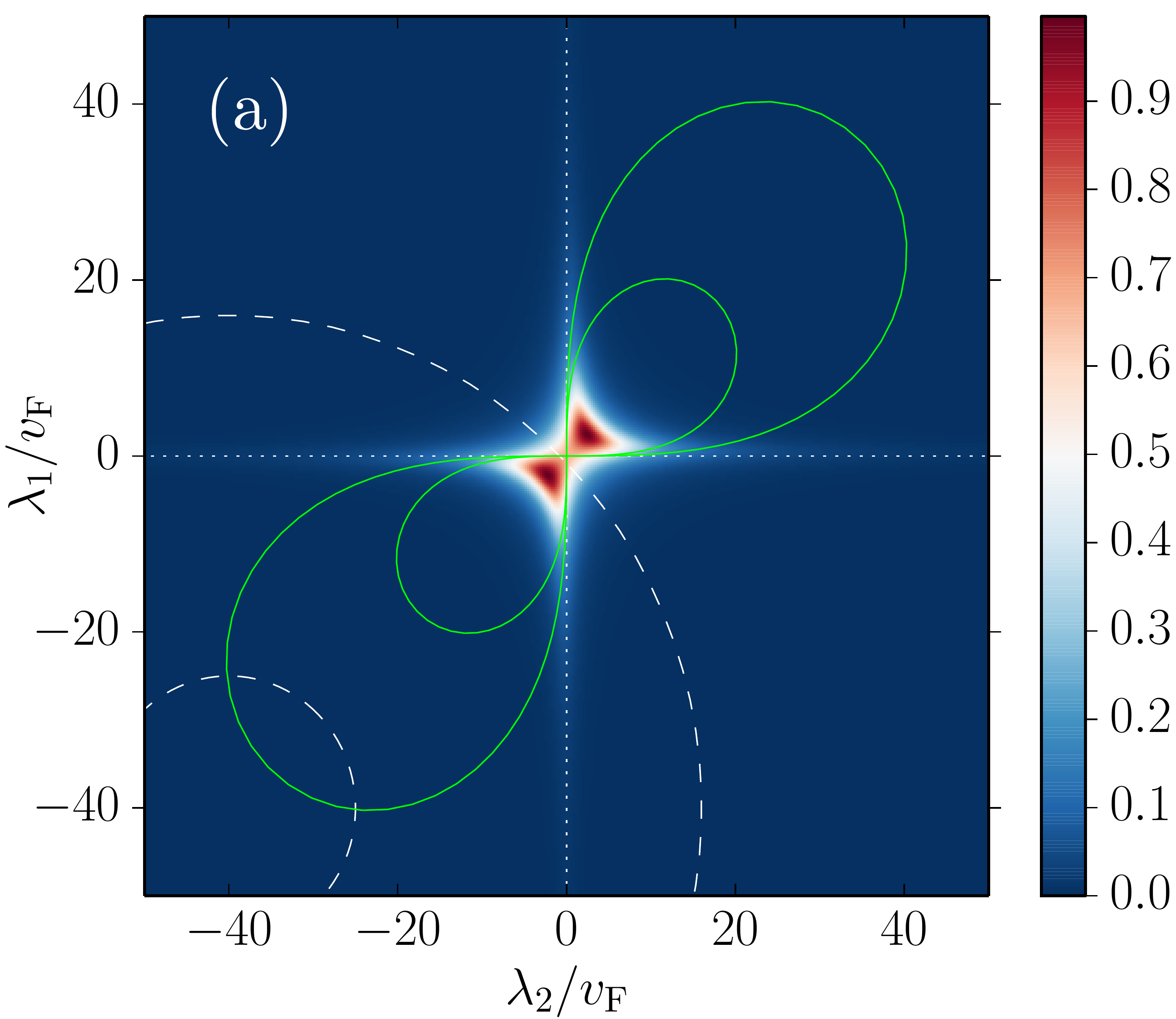}
\includegraphics[width=1.0\linewidth]{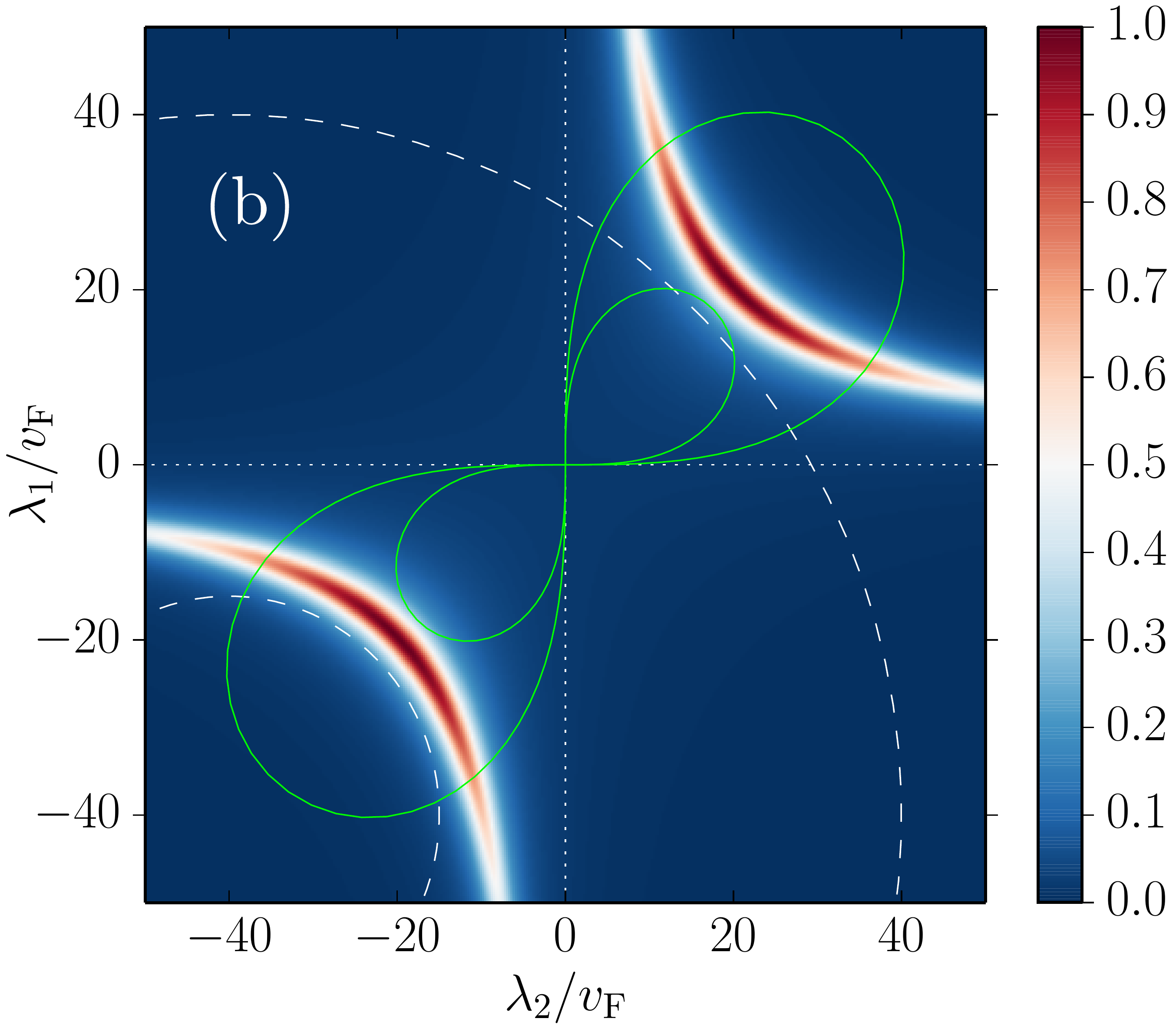}
\caption{(Color online) Contour plot of the transmission probability ($T=|t|^2$) in the $\lambda_{1}-\lambda_{2}$ plane, for transport through the \ff bound states in the \cdwqp setup, 
where $L=\xi$ for the panel (a) and $L=3\xi$ for the panel (b). All other parameters are $E=0$, $\theta=\pi/2$, $\Delta_0=E_{\rm F}/10^3$, and $k_l\simeq k_{\rm F}$. 
Increasing the wire length, and hence the separation between the \ffs, causes the two resonances to move apart from each other in the $\lambda_1$-$\lambda_2$ plane.
{The two circular pumping contours in the panel (a) correspond to $\phi=\pi/4$, $\lambda_0=-40v_{\rm F}$, $P_s=15 v_{\rm F}$ 
(smaller circle) and $P_s=40\sqrt{2}v_{\rm F}$ (bigger circle). 
For the two lemniscate contours we chose $\Theta=\pi/4$, $P_s=25 v_{\rm F}$ and $P_s=50 v_{\rm F}$.}  
In the panel (b), for the circular contours we have $\lambda_0=-40v_{\rm F}$, $P_s=25 v_{\rm F}$ and $80v_{\rm F}$ respectively, while for the lemniscate contours $\Theta=\pi/4$,  
$P_s=25 v_{\rm F}$ and $50 v_{\rm F}$.
}
\label{fig2}
\end{center}
\end{figure}
%---------------------------------------------------------------------------------------------------------------

%---------------------------------------------------------------------------------------------------------------
\begin{figure}[h!]
\begin{center}
\includegraphics[width=1.0\columnwidth]{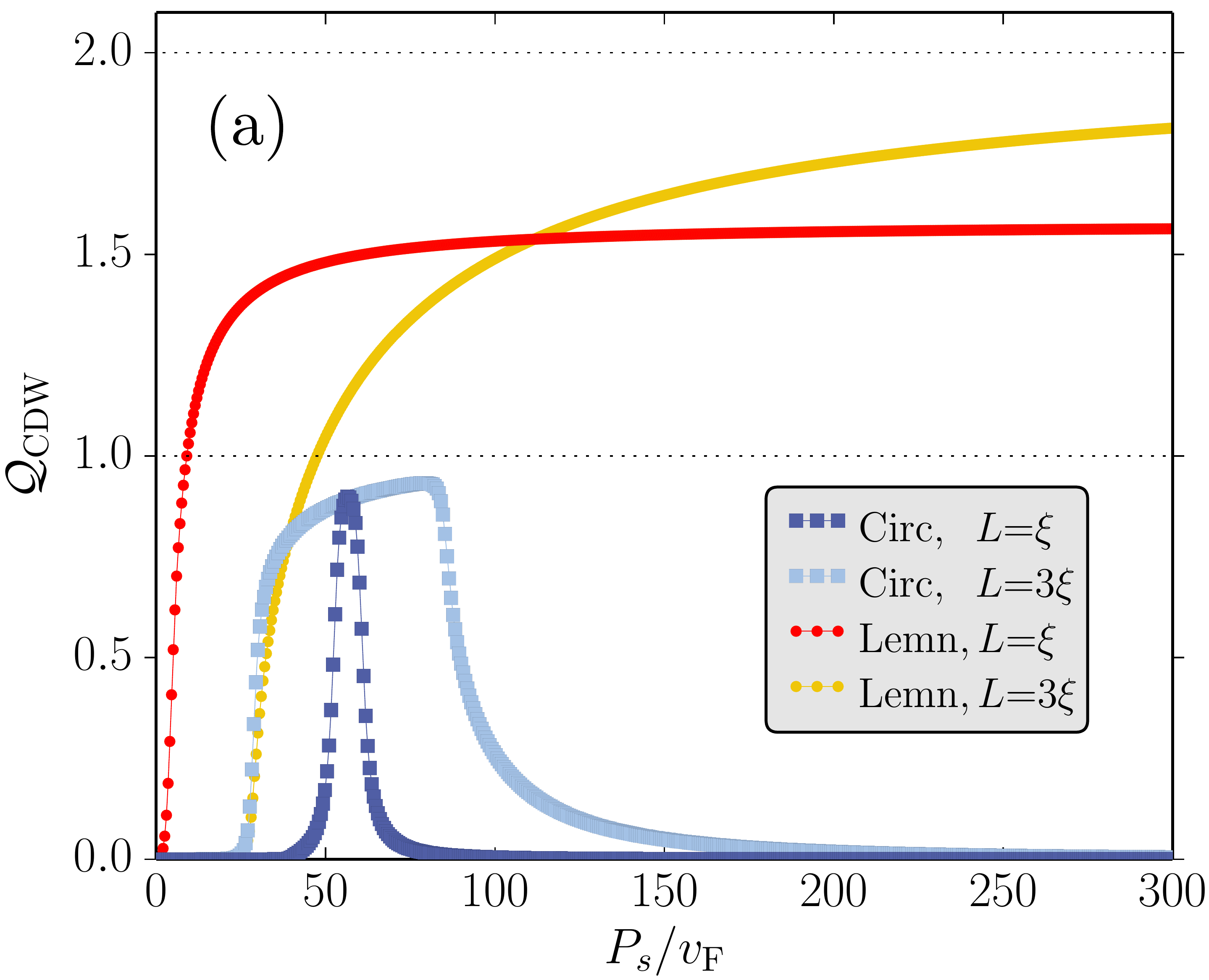}
\includegraphics[width=1.0\columnwidth]{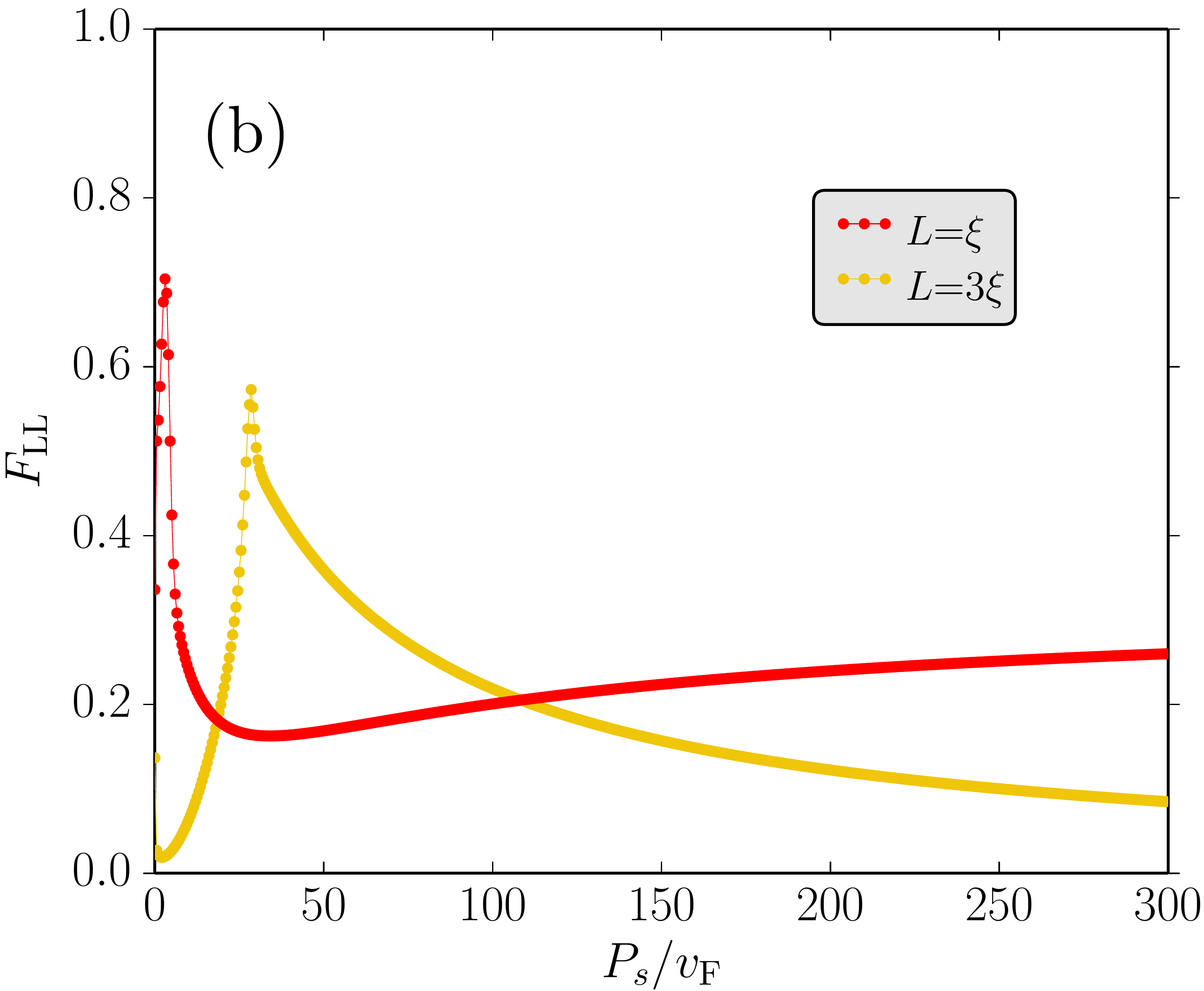}
\caption{(Color online) (a) The pumped charge $\mathcal{Q}_{\cdwqp}$ in units of the electron charge $e$ is shown as a function of the pumping strength $P_s$ 
for the \cdwqpd. 
(b) The Fano factor corresponding to the auto-correlation noise for the lemniscate curves (calculated using Eqs.~\ref{eq:deltaslcdw} and \ref{eq:deltasllcdw})
are shown as a function of the pumping strength $P_s$ for the \cdwqp at $\Theta=\pi/4$ and $\omega=\Delta_0/50$  (the circular paths give zero auto-correlator noise).
All other parameters are identical to those used in Fig.~\ref{fig2}. }
\label{fig3}
\end{center}
\end{figure}
%---------------------------------------------------------------------------------------------------------------
%
%---------------------------------------------------------------------------------------------------------------
The pumped charge for the \cdwqp is obtained by using Eq.~(\ref{pump1cdw}) with $\lambda_1$ and $\lambda_2$ as the two pumping parameters. 
They can be varied by periodically varying additional gate voltages~\cite{giblin,Giazottoetal} (not shown in Fig.~\ref{fig1}). 
The localization length of the \ffs in the \cdwqp is determined by the single energy gap, $\xi(E)\simeq \hbar v_{F}/\sqrt{\Delta_{0}^2-E^2}$.
For this geometry, we choose the \ff bound state energy at $E=0$ for $\theta=\pi/2$ such that the two \ff bound states at the two ends of the 
\nw are formally degenerate~\cite{gan12}. 
Still, they have a finite overlap and present an energy splitting due to the finite length of the \nwd.

Using Eq.~(\ref{pump1cdw}) we obtain the pumped charge through the \ff bound states in the \cdwqp case for various parameters of the system. 
In Fig.~\ref{fig3}(a) we show the pumped charge through the \ff bound states in units of electron charge $e$ as a function of the strengths of the pump parameters 
$P_s$ for the \cdwqp geometry. 
In some optimal regimes (defined below) we find that for circular ($\phi=\pi/4$) contours, described by the pumping parameters $\lambda_1=\lambda_0+P_s \cos(\omega t - \phi)$  
and $\lambda_2=\lambda_0+P_s \cos(\omega t + \phi)$, the  pumped charge can reach ${\cal Q} \sim e$ while the pumping strength $P_s$ is varied. 
On the other hand we find that the pumped charge through the \ff bound states asymptotically approaches the quantized value $2e$ in the limit of large pumping strengths
for lemniscate contours ($\Theta=\pi/4$), defined as $\lambda_1=P_s (\cos\Theta\cos\omega t - \sin\Theta\sin\omega t\cos\omega t)/(1 + \sin\omega t)^2$ and 
$\lambda_2=P_s (\cos\Theta\cos\omega t + \sin\Theta\sin\omega t\cos\omega t)/(1 + \sin\omega t)^2$ as before.

To analyze the shot noise for the \cdwqpd, we calculate the Fano factor $F_{\rm LL}^{\cdwqpd}=\Delta\mathcal{S}^{\text{pump}}_{\rm LL}/\Delta\mathcal{S}_{\rm L}^{\text{pump,P}}$ 
for the auto-correlator. The Fano factor as defined here, is a measure specific to the additional noise generated by quantum pumping.
Fig.~\ref{fig3}(b) we show $F_{\rm LL}^{\cdwqpd}$ as a function of the strength of the pump parameters $P_s$ 
for $\Theta=\pi/4$. We find that in the 
limit of large pumping strengths when the pumped charge asymptotically approaches the quantized value, the auto-correlator vanishes, 
signifying optimal pumping. 
For $\phi=\pi/4$, the auto-correlator trivially vanishes, which can be seen from Eqs.~(\ref{eq:deltaslcdw}--\ref{eq:deltasllcdw}).

To understand the behavior of the pumped charge as a function of the pumping strength $P_s$ we investigate the transmission probability ($T=|t|^2$) through the 
\ff bound states in the $\lambda_{1}-\lambda_{2}$ plane. 
In Fig.~\ref{fig2}, we plotted $T(\lambda_1,\lambda_2)$ together with 
different possible pumping contours. 
$T(\lambda_1,\lambda_2)$  exhibits transmission resonance {\it lines}, containing a resonance point $T=1$ 
and presenting a mirror symmetric behavior 
about the $\lambda_1$=$\lambda_2$ and about the $\lambda_1$=-$\lambda_2$ axes, as is apparent from the plots. 

We find that the typical pumping contours can be generically classified into three categories. Those which {\textsf{(a)}} do not enclose any $T=1$ resonance point
through the \ff bound states (e.g. smaller circles in Fig.~\ref{fig2}), {\textsf{(b)}} enclose only one resonance (e.g. bigger circles in Fig.~\ref{fig2}) and
finally {\textsf{(c)}} enclose both the resonances related to  the \ff bound states [e.g. both lemniscate contours in Fig.~\ref{fig2}(a)]. 
Further, when a contour encloses spectral weight from both resonances, the relative integration direction around the two singular points plays an important role. 
Namely, when the two resonances are enclosed in a path with the same orientation, then the two contributions have opposite sign and tend to cancel each other.
This is why for the circular paths, which do enclose the two resonances within same contour orientation, in the limit of very large integration contours when all the 
spectral weight is collected the total pumped charge tends to zero [see Figs.~\ref{fig3}(a) and \ref{fig5}(a)].
On the opposite, when the two resonances are enclosed within opposite integration orientations, the two contributions for the pumped charge sum up.
This is exactly the reason that motivates the choice of lemniscate contours.
Looking indeed again at Figs.~\ref{fig3}(a) and \ref{fig5}(a), we see that $\cal Q$ for the lemniscate contours does {\it not} go to zero for large $P_s$ and, instead, 
increases monotonically.

Another interesting observation concerns the role played by the wire length. 
The behavior of transmission probability ($T=|t|^2$) for two different wire lengths, $L=\xi$ and $L=3\xi$, is shown in Fig.~\ref{fig2}(a)-(b) respectively. 
The main observation is that for longer wires 
the energy splitting between the two \ff states is reduced and the separation of the two resonances in the $\lambda_1-\lambda_2$ plane correspondingly increases~\cite{lev}. 
In turn, this implies that the spectral weights from the two resonances overlap less and it is possible to collect full contribution from both resonances (summing up in different ways for circular and lemniscate contours).
This is why, in the pumped charge plots of Fig~\ref{fig3}(a), the light blue curve ($L$=$3\xi$) reaches a higher value than the blue curve ($L$=$\xi$), close to ${\cal Q}= e$,  before decreasing again when the second resonance starts contributing.
For the same reason, the yellow curve ($L$=$3\xi$) referring to the lemniscate contour in the same plot asymptotically tends to a higher value than the red curve ($L$=$\xi$), and in the limit of large resonance separation that value becomes ${\cal Q}=2e$.
We also note that for the same pair of realizations, the corresponding Fano factor of Fig.~\ref{fig3}(b) tends to finite constant for $L$=$\xi$ but tends to zero (or a much smaller constant) for $L$=$3\xi$, 
implying noiseless pumping when the pumped charge is quantized (${\cal Q}=2e$). %}

%=======================================================================
\subsection{\rnwqpd}
%=======================================================================

%-----------------------------------------------------------------------------------------------------------------
\color{black}
\begin{figure}[h!]
\begin{center}
\includegraphics[width=1.0\linewidth, clip=true]{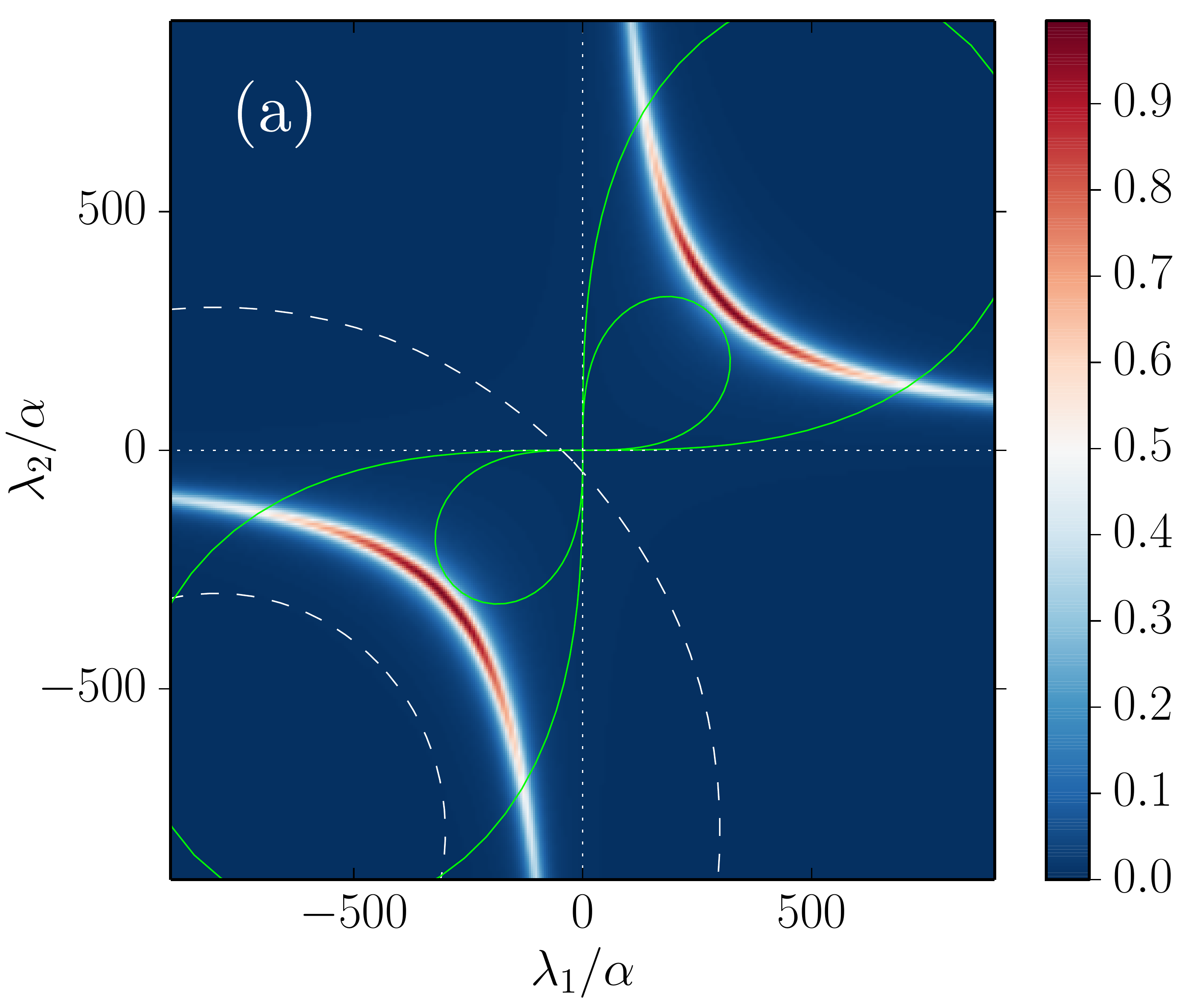}
\includegraphics[width=1.0\linewidth, clip=true]{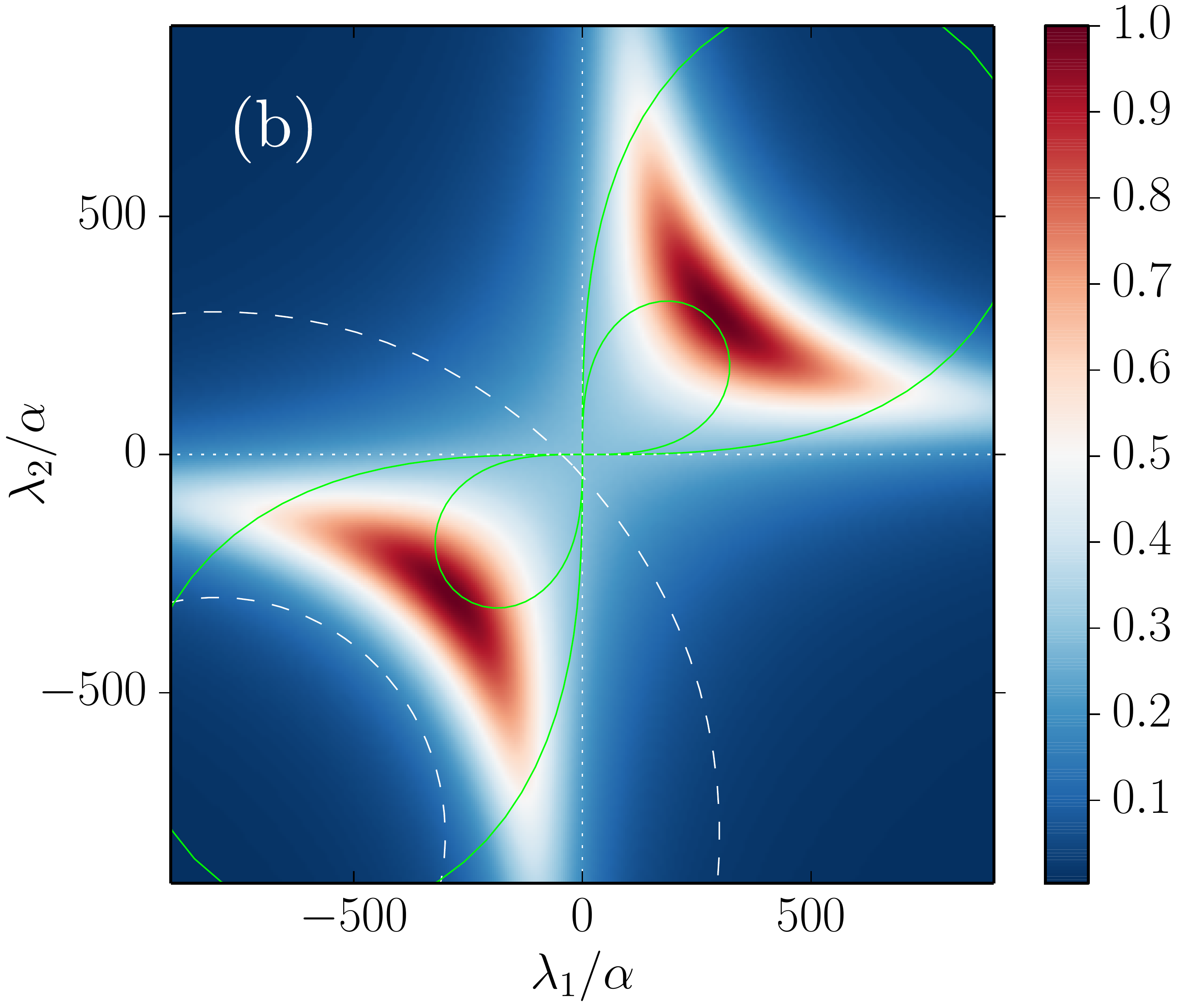}
\caption{(Color online) Contour plot of the transmission probability ($T=|t_{\uparrow\uparrow}|^2+|t_{\uparrow\downarrow}|^2$) through the \ff bound states 
in the $\lambda_{1}-\lambda_{2}$ plane for the \rnwqpd. 
The plotted results refer to $E=0$, $\theta=\pi$, 
$\Delta_z=\alpha^2/5$, $\Delta_n=\alpha^2/10$, $L=5\xi$,  $k_l\simeq 10 k_{\rm so}$ for the panel (a) and $k_l\simeq 100 k_{\rm so}$ for the panel (b). 
Different pumping contours are also shown. 
The two circular contours in both (a) and (b) correspond to $\phi=\pi/4$, $\lambda_0=-8\cdot10^2\alpha$, $P_s=5\cdot10^2\alpha$ for the smaller circle and 
$P_s=1.1\cdot10^3\alpha$ for the bigger circle, respectively. 
Similarly, for the lemniscate contours we chose $\Theta=\pi/4$, $P_s=4\cdot10^2\alpha$ and $P_s=1.2\cdot10^3\alpha$.} 
\label{fig4}
\end{center}
\end{figure}
%---------------------------------------------------------------------------------------------------------------
%

We obtain the pumped charge for the \rnwqp using Eq.~(\ref{pump1}) with $\lambda_1$ and $\lambda_2$ as the two pumping parameters 
like in the \cdwqp geometry. 
The localization length of the \ffs is physically set by the two energy gaps, $\xi_{z}(E)\simeq \hbar\alpha/\sqrt{\Delta_{z}^2-E^2}$ and $\xi_{n}(E)\simeq \hbar\alpha/\sqrt{\Delta_{n}^2-E^2}$.
In our numerical analysis we choose the \ff bound state energy at $E=0$ for $\theta=\pi$ such that the two \ff bound states at the two ends 
of the \nw are degenerate~\cite{die13}. 
Again, the finite wire length induces a finite overlap between the two \ffs and an energy splitting between the two levels.

\begin{figure}[h!]
\begin{center}
\includegraphics[width=1.0\columnwidth]{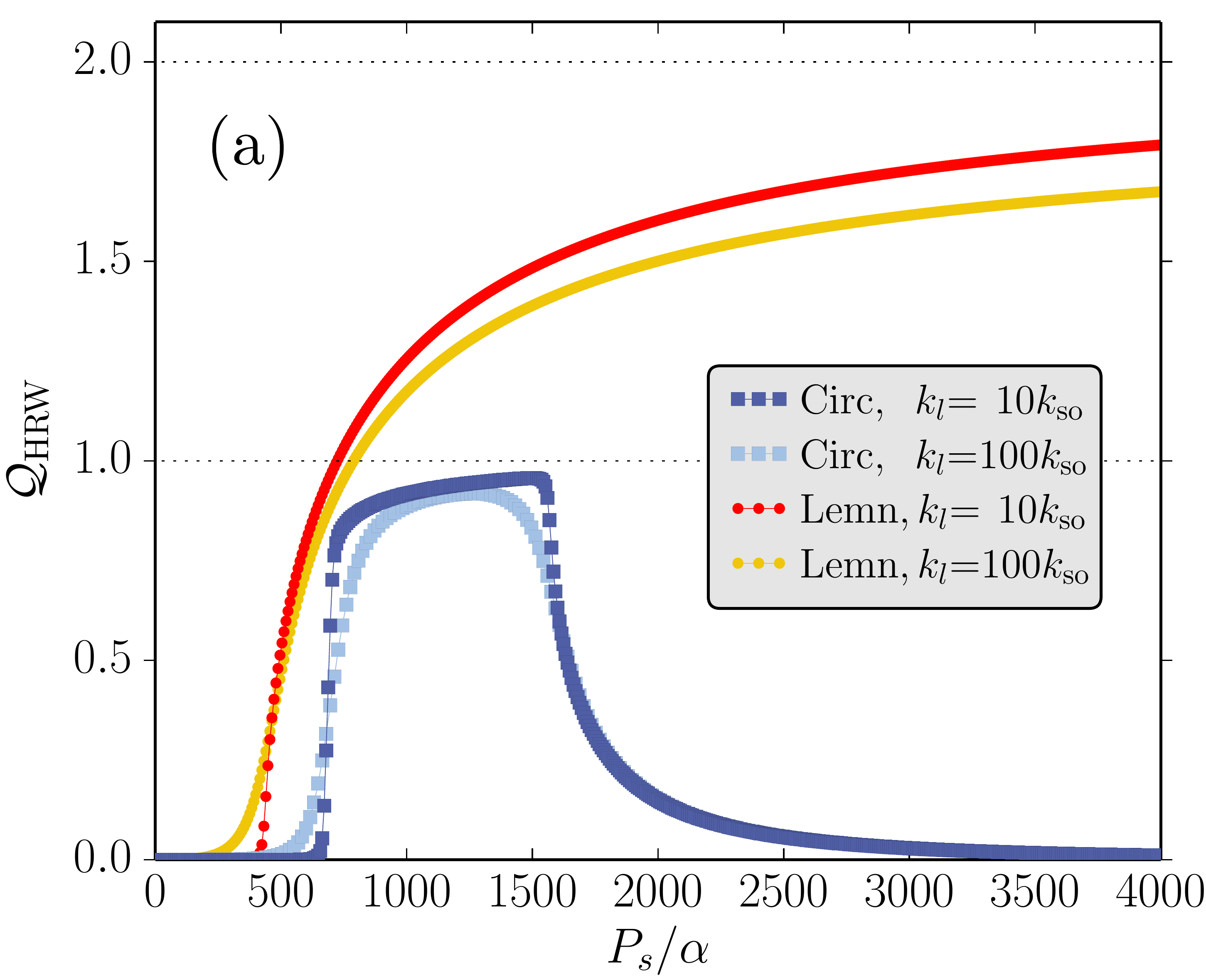}
\includegraphics[width=1.0\columnwidth]{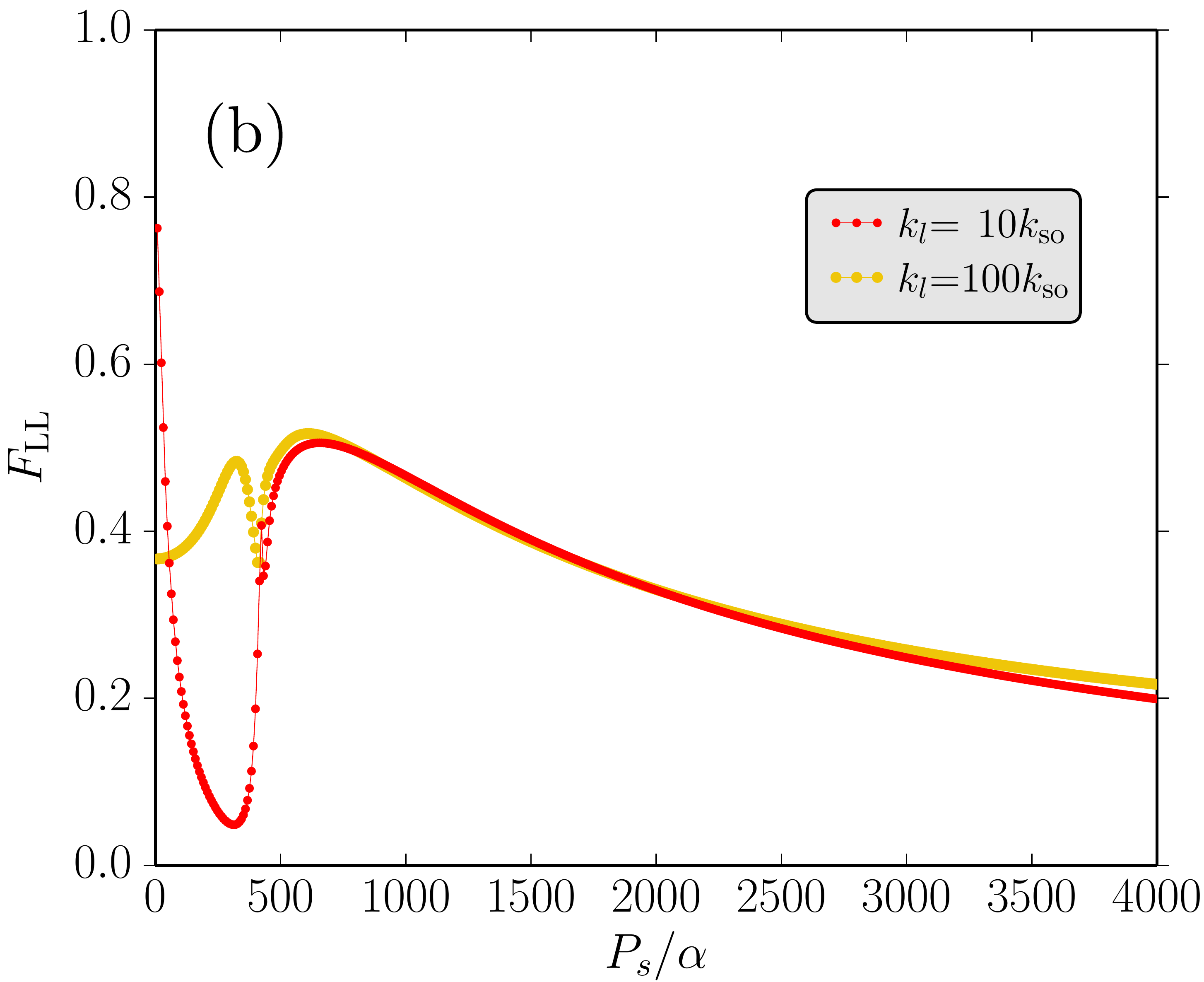}
\caption{(Color online) (a) The pumped charge $\mathcal{Q}_{\rnwqp}$ in units of the electron charge $e$ for the \rnwqp is shown as a function of the pumping strength $P_s$. 
As in Fig.~\ref{fig4}, we are plotting the results for two different values of the chemical potential in the leads$-$ $k_l=10k_{\rm so}$ and $k_l=100k_{\rm so}$. 
Here $L=5\xi$ and the pumping rate $\omega=\Delta_n/10$. 
All other parameters are the same as those used in Fig.~\ref{fig4}.
(b) The Fano factor $F_{\rm LL}$ of the auto-correlator, associated to the same lemniscate contours considered in panel (a), 
shown as a function of the pumping strength $P_s$ at $\Theta=\pi/4$ (the circular paths give zero auto-correlator noise).
}
\label{fig5}
\end{center}
\end{figure}

In Fig.~\ref{fig5}(a), we show the behavior of the pumped charge in units of electron charge $e$ as a function of the pumping strength $P_s$ for the 
\rnwqp geometry considering circular ($\phi=\pi/4$) and lemniscate ($\Theta=\pi/4$) contours. 
Similarly to the \cdwqp case, for the \rnwqp  we find that in some regimes circular contours yield large finite pumped charge (${\cal Q}\sim e$) 
at appropriate values of the pumping strength $P_s$. 
Analogously, the pumped charge for lemniscate contours can asymptotically approach
the quantized value of $2e$ in the limit of large pumping strengths. 

%
%---------------------------------------------------------------------------------------------------------------
%---------------------------------------------------------------------------------------------------------------
%

Following the analysis of the \cdwqpd, we calculate the Fano factor for the \rnwqpd, 
$F_{\rm LL}^{\rnwqpd}=\Delta\mathcal{S}^{\text{pump}}_{\rm LL}/\Delta\mathcal{S}_{\rm L}^{\text{pump,P}}$
[see Eqs.~(\ref{eq:deltasl}--\ref{eq:deltasll}) for further details]. 
In Fig.~\ref{fig5}(b) we show $F_{\rm LL}^{\rnwqpd}$ as a function of the strengths of the pump parameters $P_s$ for $\Theta=\pi/4$.
Like in the \cdwqp case, in the \rnwqp  the auto-correlator vanishes when the pumped charge approaches the quantized value, signifying optimal pumping. %}}
Additionally, the auto-correlator also vanishes for $\phi=\pi/4$ [see Eqs.~(\ref{eq:deltasl}--\ref{eq:deltasll})].

%------------------------------------------------------------------------------------------------------------------

The quantity under investigation in this \rnwqp geometry is the total transmission probability 
($T=|t_{\uparrow\uparrow}|^2+|t_{\uparrow\downarrow}|^2$) through the \ff bound states, considered to vary as a function of $\lambda_{1}$ and $\lambda_{2}$.
In Fig.~\ref{fig4} we show the behavior of the transmission probability in the $\lambda_{1}-\lambda_{2}$ plane along with different pumping contours as classified earlier for the \cdwqp case. 
In this section we are focusing on the long-wire limit ($L$=$5\xi$), exhibiting well-separated resonances, since we know from the previous CDW analysis that such regime provides the largest pumped charge. %}

In Figs.~\ref{fig4}(a)-(b) we show the behavior of $T$ for two different values of $k_{l}$ (i.e, two different values of $\mu_l$). 
Different $k_{l}$ values induce different wire-lead couplings~\cite{die13}.
For $k_{l}\simeq 10k_{\rm so}$, we obtain two sharp well-separated resonances due to small momentum mismatch between the wire and the leads. On the other hand, two resonances
become broadened for $k_{l}\simeq 100k_{\rm so}$ due to large momentum mismatch~\cite{lev}. 
The striking similarity between Fig.~\ref{fig2} and Fig.~\ref{fig4} originates from the $\delta$-function barriers at the two ends of the \nw in both quantum pumps. 
That is, in both cases we have a double-barrier problem with a resonant level in between.
The different parameters and the different energy scales involved in the two problems just imply that different pumping strengths would be needed to observe quantized 
pumped charge through the \ff bound states in them (compare the axes ranges in Figs.~\ref{fig2} and \ref{fig4}). 
We show the corresponding behavior of the pumped charge for the \rnwqp in Fig~\ref{fig5}(a), for the two considered $k_{l}$ values. 
For circular contours, we obtain almost quantized ($\mathcal{Q}\sim e$) value of pumped charge over a finite range of pumping strengths ($500\alpha <P_{s}<1500 \alpha$). 
Over this range of $P_{s}$, circular contours enclose only one of the resonances, resulting in $\mathcal{Q}\sim e$. 

The lemniscate contours give a similar phenomenology: both values of $k_{l}$ produce large pumped charge, but the case $k_{l}\simeq 10k_{\rm so}$ translates into a charge value closer to ${\cal Q}=2e$.
Finally, in Fig.~\ref{fig5}(b) we plot the Fano factor $F_{\rm LL}$ of the auto-correlator, which exhibits a singular behavior in correspondence of the $P_s$ value for which the pumped charge becomes macroscopic, that is, in correspondence of the point in the $\lambda_1$-$\lambda_2$ plane where the contour crosses the resonance point.
The smoothness of the $F_{\rm LL}$ curves changes with $k_l$.
Finally, in the limit of large $P_s$ where the lemniscate contours enclose both resonances and their entire weight, with a pumped charge ${\cal Q}\simeq2e$, the noise decreases and $F_{\rm LL}$ slowly tends to zero.

%----------------------------------------------------------------------
\section{Summary and Conclusions} \label{sec:con}
%----------------------------------------------------------------------
We have studied adiabatic quantum pumping in two different setups which support zero-energy \ff bound states in the fully gapped system. 
One is spin degenerate and based on a charge-density-wave-modulated wire (\cdwqpd), while the other one is a spinful system 
based on a Rashba nanowire in the presence of an oscillating magnetic field (\rnwqpd).
The presence of \ffs at the two ends of these wires dramatically changes the calculated pumped charge in the adiabatic regime. 
In both these fully gapped systems the charge is pumped from one lead to the other via the zero energy \ff bound states. 
We find that for certain type of pumping contours (lemniscate contours) it is possible to observe quantized pumping in the limit of large pumping strengths, 
where two units of charge are pumped in every pumping cycle.
We also calculate the shot noise for both these pumps and find that it vanishes in the regime of quantized pump charge, indicating optimal pumping. 
In both cases we find that our numerical results are in excellent agreement with the bilinear response limit~\cite{bro98} for small pumping strengths. 

Another possible pair of pumping parameters for a semi-infinite \nw could  be the strength of a single $\delta$-function barrier and the geometrical angle $\theta$.
In this situation too, one can obtain a transmission resonance through the zero-energy \ff bound states in the pumping parameter space. 
Consequently, choosing appropriate pumping contours that enclose the transmission resonances one can obtain quantized pumped charge. 

The behavior of quantized pumped charge has also been reported for many other systems where one studies quantum pumping through nanostructures. Integer pumped 
charge has been shown for pumping through open quantum dots~\cite{ewaa1,aleiner,saha} as well as through Luttinger liquids~\cite{sharma2,das2005,amit,sahadas}. 
In more recent times, similar behaviour of pumped charge has been predicted in superconducting wires with Majorana fermions~\cite{gib13}.

As far as the practical realization of the quantum pumping setups is concerned, it should be possible to fabricate such setups with the currently available experimental techniques. 
For instance, \rnwqp can be fabricated using \insbd, with $g\simeq50$ and \so energies of the order of $50~\mu$eV~\cite{VMourik} satisfying
the requirement of strong-\so regime considered in the above theoretical calculations. Hence considering typical numbers for the magnetic field intensity generated 
by the nearby nanomagnets, $B_n\simeq50~$mT~\cite{Karmakar_Nanomagnets}, one can obtain a Zeeman coupling of the order $\Delta_n\simeq 40~\mu$eV, corresponding 
to a frequency $\simeq60~$GHz. It is convenient to choose similar values also for the uniform field $\bm B$, so that the two gap values are compatible~\cite{die13}. 
 The pumping parameters, which here are the strengths of the two tunnel barriers ($\delta$-functions in our case),  
could correspond to the electrostatic potential of \textit{thin finger gates}~\cite{VMourik}.
The time period of the oscillating gate voltages $T\simeq3~$ns
is larger than the dwell time of the electrons inside the \nw $\tau_{\rm dwell}\simeq 30~$ps, hence satisfying the adiabatic condition for the quantum pump.
The pumped current through the \ff bound states should be in the range of $\simeq10-15$~pA 
and possibly be measurable in experiment with a \nw of length $L\simeq1~\mu$m.

\vspace{-0.5cm}

\acknowledgments{This work is supported by the Swiss NSF and NCCR QSIT.}

\bibliography{Qpump_FF_ref} 

\end{document}